\documentclass[amsmath,amssymb,A4]{revtex4}

\usepackage{txfonts}
\usepackage{color}
\usepackage{amssymb}
\usepackage{graphicx}
\usepackage{color}
\usepackage{dcolumn}
\usepackage{bm}
\usepackage{multirow}

\begin{document}

\newcommand{\kfa}{KFe$_2$As$_2$}
\newcommand{\bkfa}{Ba$_{1-x}$K$_x$Fe$_2$As$_2$}
\newcommand{\tc}{$T_c$}
\newcommand{\hc}{$H_{c2}(T)$}
\newcommand{\ttt}{$T_0$}

\title{Fermi Surface of \kfa\ from Quantum Oscillations in Magnetostriction}

\author{D. A. Zocco}
\email[E-mail: ]{diego.zocco@kit.edu}
\author{K. Grube}
\author{F. Eilers}
\author{T. Wolf}
\author{H. v.\,L\"{o}hneysen}
\affiliation{Institute of Solid State Physics (IFP), Karlsruhe Institute of Technology, D-76021 Karlsruhe, Germany.}

\date{October 19, 2013}

\begin{abstract}
We present a study of the Fermi surface of \kfa\ single crystals. Quantum oscillations were observed in magnetostriction measured down to 50\,mK and in magnetic fields $\mu_0 H$ up to 14\,T. For $H$\,$\parallel$\,$c$, the calculated effective masses are in agreement with recent de Haas--van Alphen and ARPES experiments, showing enhanced values with respect to the ones obtained from previous band calculations. For $H$\,$\parallel$\,$a$, we observed a small orbit at a cyclotron frequency of 64\,T, characterized by an effective mass of $\sim$\,0.8\,$m_e$, supporting the presence of a three-dimensional pocket at the Z-point.
\end{abstract}

\maketitle

\section{Introduction}\label{intro}
The multiband character of the Fermi surface (FS) of \kfa\ deeply affects the properties of the normal and superconducting states. In a recent study \cite{zocco13a}, we showed that multiband effects are manifest in the thermal-expansion coefficient $\alpha / T$, and in the upper-critical field $H_{c2}(T)$. In addition, $H_{c2}$ is strongly affected by Pauli-limiting when fields are applied parallel to the FeAs layers \cite{zocco13a,burger13a}. This compound belongs to the \bkfa\ series, in which the superconducting state reaches a maximum \tc\ of 38\,K for $x$\,=\,0.4 \cite{rotter08a}. Further hole-doping with K suppresses \tc\ to a minimum of 3.4\,K for $x$\,=\,1. It has long been argued that the symmetry of the superconducting order parameter varies across the series. In the optimally doped compound, the superconducting state is fully gapped and exhibits $s_{\pm}$-wave symmetry \cite{ding08a}. For \kfa, on the other hand, evidence for multi-gap nodal $s$-wave superconductivity has been found in nuclear quadrupole resonance \cite{fukazawa09a} and angle-resolved photoemission spectroscopy (ARPES) \cite{okazaki12a} experiments, while recent work report a possible $d$-wave pairing mechanism \cite{tafti13a}. The latter suggests that a Lifshitz transition or a phase transition with broken symmetry could take place at intermediate K concentrations.

In order to understand the superconducting properties of these compounds, the structure of the FS has been widely studied. In the optimally doped \bkfa, the FS is derived from multiple Fe-3$d$ orbitals, composed of three concentric quasi-two-dimensional hole cylinders located at the Brillouin-zone (BZ) center ($\Gamma$-point), and electron cylinders located at the zone corner. As the chemical potential changes with increasing K concentration, the corner cylinders transform into hole Fermi sheets \cite{sato09a}. The FS of \kfa\ is formed by quasi-two-dimensional hole cylinders aligned with the $k_z$-axis: three ($\alpha$, $\zeta$ and $\beta$) centered at the $\Gamma$-point, and hole cylinders $\epsilon$ near the X-point \cite{terashima10a,hashimoto10a,okazaki12a,yoshida12a,terashima13a}. Additionally, small three-dimensional hole pockets were predicted theoretically to exist at the Z-point \cite{terashima10a,hashimoto10a}. To what extent these multiple bands affect the structure of the superconducting gap in \kfa\ is a subject of extensive study \cite{hardy13b}.

In this work, we studied the Fermi surface of \kfa\ via quantum oscillations (QOs) in magnetostriction measurements down to 50\,mK. In the present manuscript we focus in the normal conducting state, from the upper critical field up to a maximum applied field of 14\,T ($H_{c2}^{c}$\,=\,1.5\,T, $H_{c2}^{ab}$\,=\,4.8\,T at $T$\,$\rightarrow$\,0). A detailed study of the superconducting properties of \kfa\ below \hc\ can be found in Ref.~\cite{zocco13a} and in the references therein.

\section{Experimental methods}\label{exp}
Single crystals of \kfa\ were grown from self flux as described in \cite{zocco13a}. Millimeter-size platelike samples with typical residual-resistivity ratios RRR\,=\,$\rho$(300\,K)/$\rho$(4\,K)\,$\sim$\,1000 were obtained. The experiments were carried out in a home-built CuBe capacitive dilatometer for measurements of thermal expansion and magnetostriction, parallel and perpendicular to the applied magnetic field. For a typical sample length of 1\,mm, the resolution amounts to $\Delta L / L$\,$>$\,10$^{-10}$. The magnetostriction coefficients are defined as $\lambda_{i}$\,=\,$L_i^{-1}\partial L_i / \partial (\mu_0 H)$, where $L_i$ is the length of the sample along the axis $i$\,=\,$a$ or $i$\,=\,$c$. Temperatures as low as 50\,mK were obtained in a commercial Oxford Kelvinox $^{3}$He-$^{4}$He dilution refrigerator equipped with a 14\,T superconducting magnet. The measurements presented here were performed on two different samples (Sample 1 and Sample 2), obtained from two different growths.

\section{Results}

\begin{figure}[h]
\begin{center}
\includegraphics[width=0.85\textwidth]{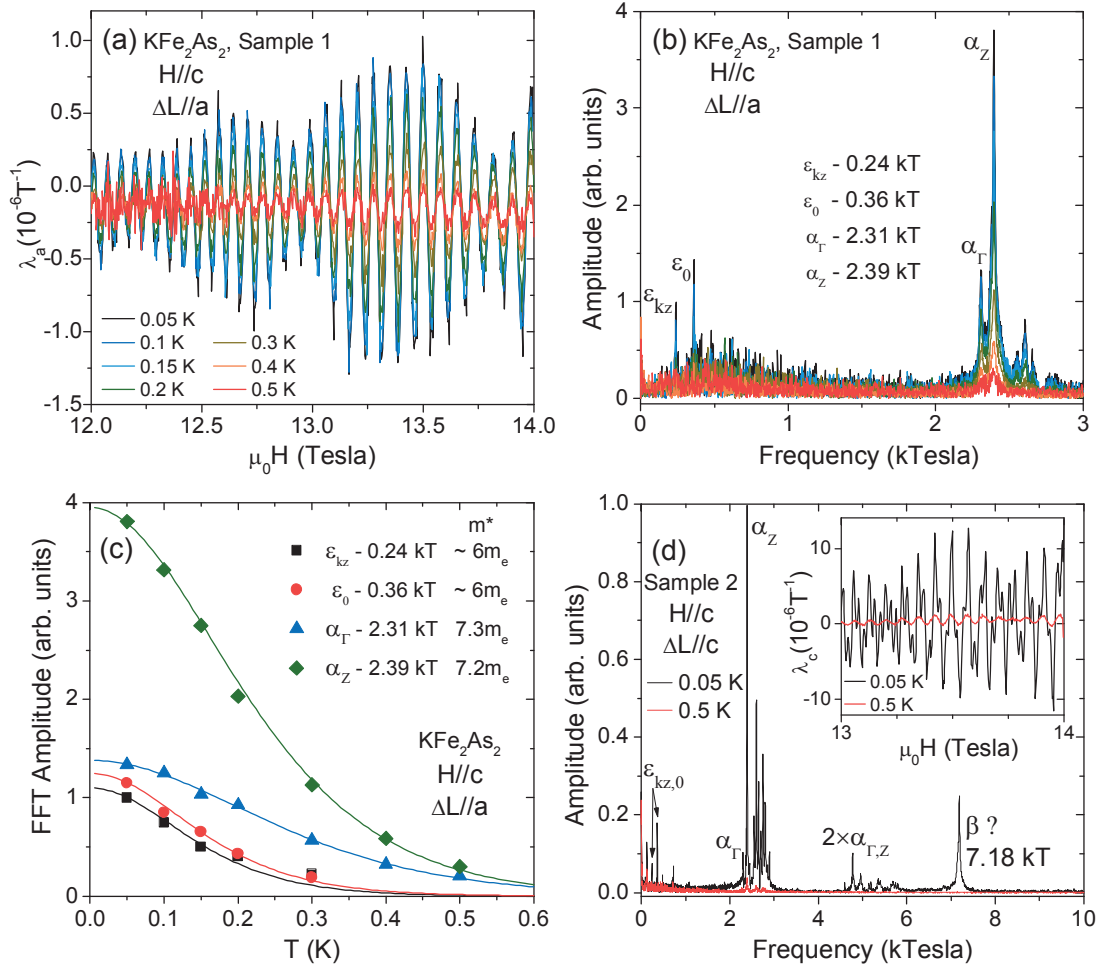}
\caption{(a) Quantum oscillations observed in magnetostriction $\lambda_a$ for magnetic field $H$\,$\parallel$\,$c$ at several temperatures (Sample 1). (b) Fourier-transform (FT) spectrum of the QOs displayed in (a), for a field window of 2--14\,T. (c) Amplitude of the peaks displayed in (b) versus temperature and the corresponding fits using Eq.~\ref{lk1}, for the calculation of the effective masses $m^*$ associated with the $\epsilon_{k_z}$, $\epsilon_{0}$, $\alpha_{\Gamma}$ and $\alpha_{\mathrm{Z}}$ Fermi sheets. For the $\epsilon$ Fermi sheets, data was obtained from a FT over a reduced field window (4--14\,T), and only low temperature data is displayed. (d) In a different sample (Sample 2), the Fourier-transform spectrum of the 50\,mK QOs revealed an additional peak at 7.18\,kT, which is completely suppressed at 0.5\,K.}
\label{fig1}
\end{center}
\end{figure}

For $H$\,$\parallel$\,$c$, quantum oscillations (QOs) in the magnetostriction $\lambda_a$ ($L$\,$\parallel$\,$a$) were observed above $H_{c2}^{c}$\,=\,1.5\,T up to 14\,T. Fig.\,\ref{fig1}(a) displays the high-field region for temperatures 0.05\,K$\leq$\,$T$\,$\leq$\,0.5\,K (Sample 1). A clear beat pattern is observed, damped progressively with increasing temperature. In fact, a Fourier-transform (FT) analysis of the QOs (periodic in 1/$H$) reveals two pairs of sharp peaks with similar cyclotron frequency values ($F$) [Fig.\,\ref{fig1}(b)], which correspond to the low and high components of the $\alpha$ and $\epsilon$ cylinders previously determined from de Haas--van Alphen (dHvA) measurements \cite{terashima10a,terashima13a}. Additional peaks are observed at 2.55, 2.61 and 2.66\,kT, which we could not identify with the cyclotron orbits of the $\zeta$ and $\beta$ cylinders, not with harmonics of the $\epsilon$ peaks.
Values of effective mass $m^*$ can be inferred from the temperature dependence of the QOs, using the Lifshitz–-Kosevich formula which gives the temperature damping factor of the amplitude $A$ of the QOs \cite{shoenberg1984}:
\begin{equation}\label{lk1}
A(T)=a_0 \frac{a_1T}{\mathrm{sinh}(a_1T)}, \qquad a_1 = \frac{2 \pi^2 k_\mathrm{B}}{e \hbar}\frac{m^*}{m_e}\frac{1}{\mu_0 H} \sim 14.7\frac{m^*}{m_e}\frac{1}{\mu_0 H}
\end{equation}
where $m_e$ is the bare electron mass. The coefficient $a_1$ can be estimated from fits to the temperature dependence of the amplitudes of either the FT peaks or the oscillations in $\lambda$. In the first case, the value of magnetic field $H$ in Eq.~\ref{lk1} is obtained from the average of the 1/$H$ window used in the FT, while for the second method, the value of $H$ corresponds to the field range at which the amplitude of the QOs in magnetostriction were determined.
\begin{figure}
\begin{center}
\includegraphics[width=0.85\textwidth]{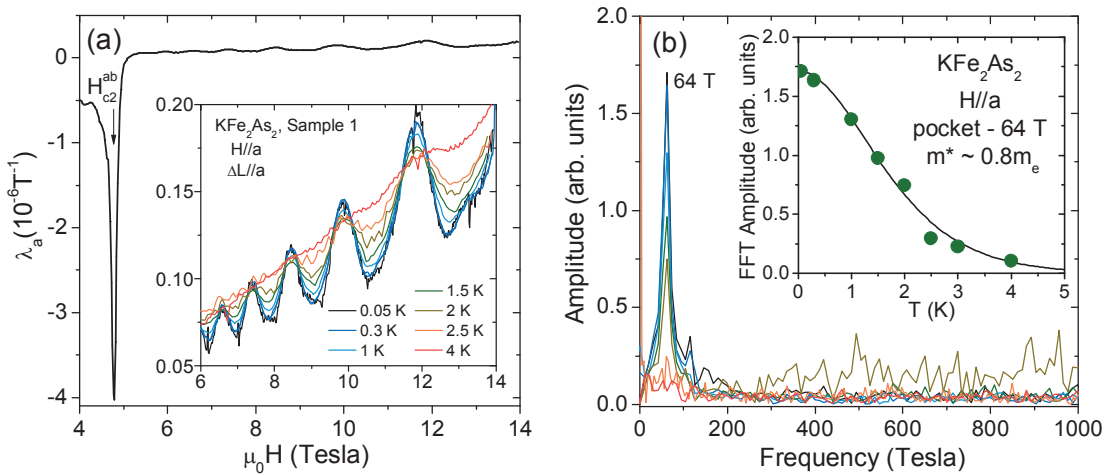}
\caption{(a) Magnetostriction $\lambda_a$ for magnetic field $H$\,$\parallel$\,$a$ at 0.05\,K. The inset displays the QOs at several temperatures observed above $H_{c2}^{ab}(T_c)$. (b) Fourier-transform spectrum of the QOs displayed in (a). The temperature dependence of the amplitude of the peak at 64\,T gives an effective mass value $m^*$\,$\sim$\,0.8\,$m_e$ (inset).}
\label{fig2}
\end{center}
\end{figure}

Fig.\,\ref{fig1}(c) shows the fits obtained for $H$\,$\parallel$\,$c$, for the $\epsilon_{k_z}$, $\epsilon_{0}$, $\alpha_{\Gamma}$ and $\alpha_{\mathrm{Z}}$ Fermi sheets. For the $\alpha_{\Gamma,\mathrm{Z}}$ oscillations ($F_{\alpha}$\,$\sim$\,2.3\,kT), it is possible to resolve the characteristic peaks in a FT window of 12--14\,T, which translates in a value of $H_{avg}$\,= $2(1/14 + 1/12)^{-1}$\,=\,12.9\,T. The effective mass values for $\alpha_{\Gamma}$ and $\alpha_{\mathrm{Z}}$ are 7.3$m_e$ and 7.2$m_e$, respectively. For the $\epsilon_{k_z,0}$ oscillations, on the other hand, that value of $H_{avg}$ results in small effective masses $\sim$\,2.5$m_e$, much smaller than the previously reported values. Therefore, larger field windows are used (4--14\,T and 6--14\,T). This allows us to resolve FT peaks up to a maximum temperature of 0.3 K, which results in effective mass values of $\sim$\,(6$\pm$1)$m_e$. Further increase of $H_{avg}$ makes the amplitudes of the FT peaks smaller and the estimation of the effective masses more inaccurate.

Quantum oscillations were also measured in Sample 2 at 0.05\,K and at 0.5\,K [Fig.\,\ref{fig1}(d)]. At 0.05\,K, the $\alpha$, 2$\alpha$ and $\epsilon$ peaks are well defined in the FT. Similarly to Sample 1, a group of additional peaks appear between 2.5 and 2.9\,kT, which we are not able to unambiguously identify with harmonics, peaks predicted by band calculations or peaks from previous dHvA measurements. A peak of larger amplitude appears at 7.18\,kT, not observed in Sample 1, which can be associated with either the third harmonic of $\alpha_{\mathrm{Z}}$ or with the $\beta$ cylinder.

For $H$\,$\parallel$\,$a$, QOs were observed above $H_{c2}^{ab}$\,=\,4.8\,T. Fig.\,\ref{fig2}(a) displays these QOs between 6 and 14\,T, for temperatures ranging from 0.05\,K to 4\,K. A low-frequency component is clearly observed at low temperatures with a characteristic frequency of 64\,T, which is nearly suppressed at 4\,K. The temperature dependence of the Fourier transform [Fig.\,\ref{fig2}(b)] results in a value of effective mass for this band of $m^*$\,$\sim$\,0.8\,$m_e$.

\begin{figure}
\begin{center}
\includegraphics[width=0.85\textwidth]{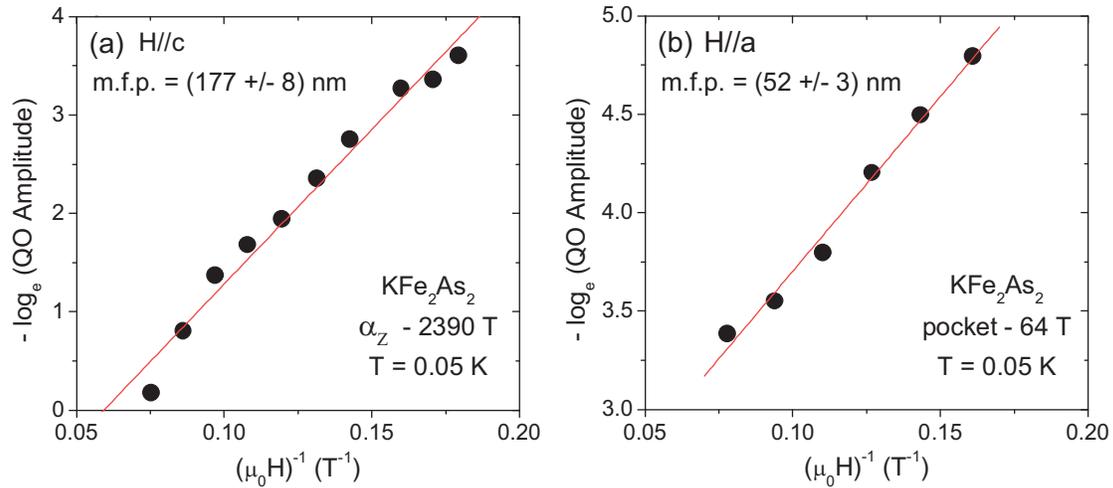}
\caption{Calculation of mean free path length from Dingle analysis: (a) $H$\,$\parallel$\,$c$, (b) $H$\,$\parallel$\,$a$.}
\label{fig3}
\end{center}
\end{figure}

Another damping effect affecting the amplitude of the quantum oscillations arises from impurities in the samples. The Dingle impurity factor can be written as
\begin{equation}\label{dingle1}
R_\mathrm{D}=\mathrm{exp}\left( -\frac{\pi m^* v_\mathrm{F}}{e \mu_0 H \ell}\right), \qquad v_\mathrm{F}=\frac{\sqrt{2e\hbar F}}{m^*}
\end{equation}
where $v_\mathrm{F}$ is the Fermi velocity, $e$ the electron charge, $\mu_0 H$ the applied magnetic field, and $\ell$ the mean free path of the quasi-particles of effective mass $m^*$. The Fermi velocity can be calculated assuming circular orbits. For $H$\,$\parallel$\,$c$, we plot in Fig.\,\ref{fig3}(a) the field dependence of the amplitude of the QOs at 50\,mK corresponding to the $F_{\alpha_{\mathrm{Z}}}$\,=\,2.39\,kT component ($m^*$\,=\,7.2\,$m_e$ and $v_\mathrm{F}$\,=\,35450\,m/s), from which we obtain $\ell$\,=\,(177$\pm$8)\,nm. For $H$\,$\parallel$\,$a$, we plot in Fig.\,\ref{fig3}(b) the amplitudes of the only visible component ($F$\,=\,64\,T, $m^*$\,=\,0.8\,$m_e$ and $v_\mathrm{F}$\,=\,63800\,m/s), giving $\ell$\,=\,(52$\pm$3)\,nm. In all cases, the amplitude values were corrected by the temperature damping factor described above. With the coherence lengths of $\xi_{ab}$\,$\sim$\,15\,nm and $\xi_c$\,$\sim$\,3\,nm \cite{terashima09a}, the ratio $\ell / \xi$\,$\sim$\,15 confirms that the samples are in the superconducting clean limit.

\section{Discussion}\label{disc}

The values of the effective mass extracted from our magnetostriction data $\lambda (H)$, and from previously reported dHvA \cite{terashima10a,terashima13a} and ARPES \cite{yoshida12a} experiments, are collected in Table~\ref{table1}. The extracted high effective masses are consistent with the enhanced Sommerfeld coefficient present in this compound. For $H$\,$\parallel$\,$c$, the values of the effective mass from $\lambda$ are in good agreement with previous experiments. Since for $H$\,$\parallel$\,$c$, $\lambda (H)$ is dominated by the average of the $\alpha_{\Gamma}$ and $\alpha_{\mathrm{Z}}$ oscillations [Fig.\,1(a)], it is also possible to calculate for the $\alpha$ band an average $m^*_{\alpha}$ directly from the amplitudes of $\lambda (H)$: at 13.5\,T, $m^*_{\alpha}$\,$\sim$\,8$m_e$, similar to the values calculated from the amplitudes of peaks determined in the Fourier-transform analysis.

\begin{table}[h]
\small
\caption{\label{table1}
Effective masses $m^*$ determined from dHvA \cite{terashima10a,terashima13a}, ARPES \cite{yoshida12a} and our magnetostriction measurements ($\lambda$). The frequencies $F$ and Fermi velocities $v_\mathrm{F}$ calculated from our data are given in the last two columns (values marked with * are taken from Ref.~\cite{terashima10a}). Right: Schematic diagrams of the first Brillouin zone for a tetragonal body-centered lattice (top) and of the FS cross section (bottom).}\vspace{5mm}
\centering
\begin{tabular}{c|c c|c c c|c c}
\cline{1-8}
    $H$-field      & {\multirow{2}{*}{label}} & {\multirow{2}{*}{$k_z$}}&  & $m^{*}/m_e$ &    &{\multirow{2}{*}{$F$ (T)}} &{\multirow{2}{*}{$v_\mathrm{F}$ (m/s)}} \\

    direction                  &        &      &dHvA&ARPES&$\lambda$&    &       \\
\cline{1-8}
{\multirow{8}{*}{$c$-axis}}&    $\alpha_{\Gamma}$      &            $\Gamma$           &   6.0   &    5.1   &            7.3         &2310&42015\\
                               & $\alpha_{\mathrm{Z}}$ &                Z              &   6.5   &    6.6   &            7.2         &2390&35450\\
                               &    $\zeta_{\Gamma}$       &            $\Gamma$           &   8.5   &   11.0   &            --          &~2890*&~40360*\\
                               & $\zeta_{\mathrm{Z}}$ &                Z              &  18.0   &    9.6   &            --          &~4400*&~23520*\\
                               &{\multirow{2}{*}{$\beta$}}&            $\Gamma$           &{\multirow{2}{*}{19}}& 16.3  &{\multirow{2}{*}{$\gtrsim$16}}&{\multirow{2}{*}{7180}}&{\multirow{2}{*}{33800}}\\
                               &                  &                Z              &         &   17.9   &                        &    &     \\
                               & $\epsilon_{kz}$     &      $2\pi /c$    &   6.0   &    5.6   &{\multirow{2}{*}{6$\pm$1}}& 240&36080\\
                               & $\epsilon_{0}$ &      0          &   7.2   &    --    &                     & 360&50870\\
\cline{1-8}
          $a$-axis         &  pocket              &        Z                      &   --    &    --    &            0.8         & 64 &63815\\
\cline{1-8}
\end{tabular}
\quad
\begin{minipage}[c]{0.23\textwidth}%
\centering
    \includegraphics[width=0.75\textwidth]{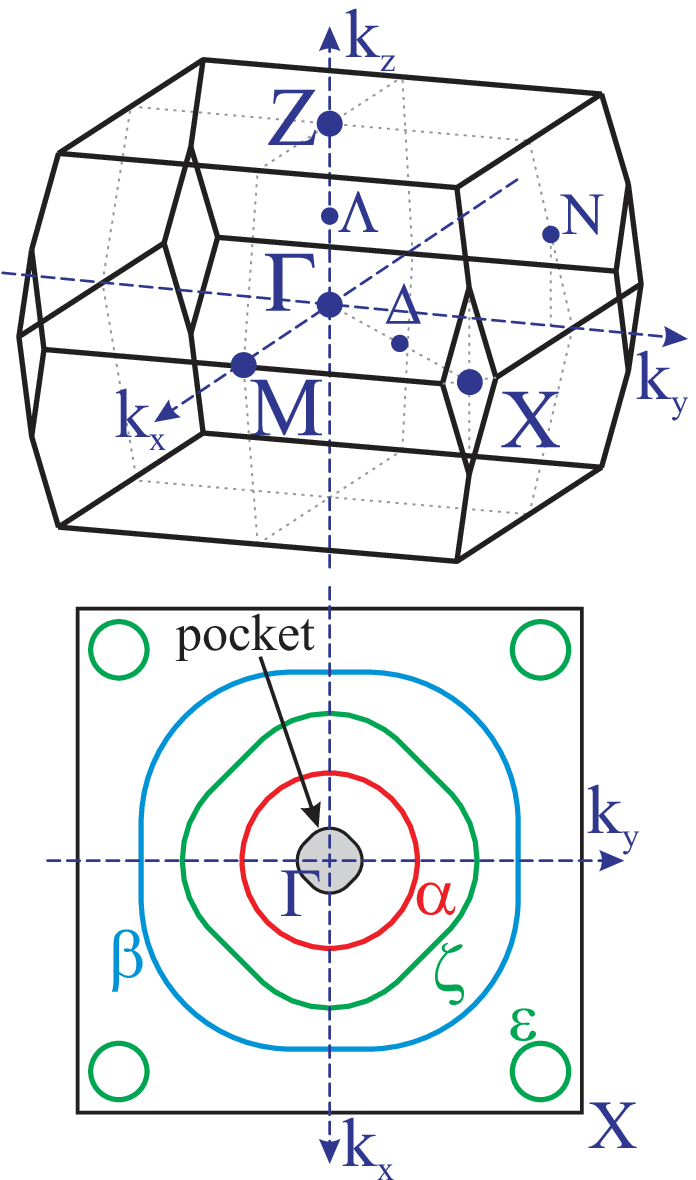}\label{figBZ}
\end{minipage}
\end{table}

In Sample 2, an additional peak appears at 7.18\,kT in the FT [Fig.\,\ref{fig1}(d)], with an amplitude larger than that of the second harmonic of $\alpha_{\Gamma,\mathrm{Z}}$. A peak associated with the $\beta$ cylinder is expected to appear at this frequency, following the estimation of the FS cross-sectional area given by ARPES \cite{yoshida12a}. Given that the amplitude of harmonics decreases exponentially with harmonic number, we can assign our 7.18\,kT peak to orbits in the $\beta$ cylinder \cite{terashima10a,terashima13a}. At 0.5\,K, this peak is completely suppressed. Unfortunately, we did not collect data for Sample 2 at intermediate temperatures, making difficult an accurate estimation of the effective mass. The amplitudes of the FT peaks at 7.18\,kT set a lower limit for $m^*/m_e$\,$\sim$\,16 for the $\beta$ cylinder, in agreement with $m^*/m_e$\,=\,16--19 recently obtained from dHvA and ARPES experiments \cite{terashima10a,yoshida12a,terashima13a}.

For $H$\,$\parallel$\,$a$ (Fig.\,\ref{fig2}), the QOs are characterized by a single low-frequency component ($F$\,=\,64\,T). They are nearly suppressed at 4\,K, resulting in an effective mass of $\sim$\,0.8$m_e$. Signatures of such small FS areas were not observed in previous dHvA or ARPES experiments \cite{sato09a,terashima10a,okazaki12a,yoshida12a,terashima13a}. Since the amplitude of QOs in magnetostriction is proportional to $d\mathrm{ln}(F)/dp_i$ \cite{shoenberg1984}, where $p_i$ is the uniaxial stress along the $i$-axis ($i$\,=\,$a,c$), small extremal FS areas can appear more prominently in magnetostriction measurements than in other techniques. Given that these oscillations correspond to out-of-plane orbits, they cannot originate from two-dimensional FSs. Their size and effective mass rather point to small three-dimensional pockets which have been predicted to exist above the zone center at the Z-point \cite{terashima10a,hashimoto10a}. The relatively small size of these pockets and of the derived effective mass suggest that these pockets do not affect significantly the electronic properties of \kfa.

\section{Conclusions}
We have studied quantum oscillations in the magnetostriction of \kfa\ single crystals. Overall, the Fermi surface inferred from our QOs is in agreement with the previously reported electronic structure of \kfa, and confirms its quasi-two-dimensional nature and the presence of strong electronic correlations. Additionally, a small orbit observed for $H$\,$\parallel$\,$a$ ($F$\,=\,64\,T, $m^*$\,=\,0.8\,$m_e$) supports the presence of a three-dimensional pocket at the Z-point.

\section{Acknowledgments}
The authors thank J. Wosnitza and C. Meingast for stimulating discussions, and R. Sch\"{a}fer and S. Zaum for help with the experiments. This work has been partially supported by the DFG through SPP1458.

\end{document}